\documentclass[11pt]{article} 

\begin{document}

\title{Propagating torsion in the Einstein frame} 
\author{Nikodem J. Pop\l awski}
\date{}

\maketitle
\begin{center}
{\it Department of Physics, Indiana University, 
727 East Third Street, Bloomington, Indiana 47405}~\footnote[0]{
E-mail: nipoplaw@indiana.edu}
\end{center}

\begin{abstract}
The Einstein--Cartan--Saa theory of torsion modifies the spacetime volume
element so that it is compatible with the connection.
The condition of connection compatibility gives constraints on torsion,
which are also necessary for the consistence of torsion, minimal coupling,
and electromagnetic gauge invariance.
To solve the problem of positivity of energy associated with the torsionic
scalar, we reformulate this theory in the Einstein conformal frame.
In the presence of the electromagnetic field, we obtain
the Hojman--Rosenbaum--Ryan--Shepley theory of propagating
torsion with a different factor in the
torsionic kinetic term.
\end{abstract}


\section{Introduction}

In the Einstein--Cartan theory, which extends general relativity
to non-symmetric connection, spin is the source of torsion~\cite{SG2}.
The equations relating torsion and spin are algebraic and torsion does not
propagate~\cite{HHKN}.
To allow such a propagation, we need a differential equation for the torsion 
tensor $S^{\rho}_{\phantom{\rho}\mu\nu}$.
This is usually achieved by modifying the Einstein--Hilbert action and
introducing a scalar field related to the torsion 
vector $S_{\nu}=S^{\mu}_{\phantom{\rho}\mu\nu}$~\cite{HRRS1, HRRS2, SG1, Sa2}.
There are also models in which the torsion vector is proportional to the
electromagnetic potential~\cite{MK, H1, H2}.

The Lagrangian density for a gravitational field can be, in principle, given
by any scalar constructed from the curvature, connection, and metric.
The standard Einstein--Hilbert Lagrangian of general relativity is linear
in the curvature scalar $R$ and torsion enters the dynamics through $R$.
Such a Lagrangian is very natural from a physical point of view since it is
linear in the second derivatives of the metric tensor.
These derivatives can be taken out of the action using the Gau\ss\,\,theorem, 
making the Lagrangian a function of the metric and its first derivatives 
only~\cite{LL2}. 
The corresponding equations of field are thus second order.
If the gravitational Lagrangian is a more general function of $R$ or other
curvature invariants, the field equations remain second order if we adopt
the Palatini variational principle according to which the 
connection and metric are {\em a priori} independent quantities and we vary 
the action with respect to both of them~\cite{Ein}.
In this approach, the connection arises from the field equations
and is not metric compatible, 
$g_{\mu\nu;\rho}\neq0$.\footnote{Except the case where the Lagrangian is 
linear in $R$ and there is no torsion.}

The Palatini variation of connection imposes four algebraic constraints on 
the matter part of the action, which are caused by invariance of the curvature 
scalar under a projective transformation of the connection~\cite{Sch, San}.
These constraints lead to inconsistencies that can be eliminated by replacing
the connection with its projective-invariant part~\cite{San}.
This procedure is equivalent to imposing the condition $S_{\mu}=0$~\cite{SL}.
In this case, we cannot associate the gradient of a scalar field with the
torsion vector and induce the propagation of torsion.
Therefore, we follow the metric variational principle and fix the connection
by assuming its metricity, $g_{\mu\nu;\rho}=0$.

In the presence of torsion, we must modify the covariant volume element so
that it remains parallel~\cite{Sa2, Sa4}.
Such a modification is possible only if $S_{\mu}$ is the 
gradient of a scalar, which gives four equations of constraint on the torsion
tensor~\cite{Sa1}.
This condition is also necessary for the compatibility of torsion and gauge
invariance of the minimally coupled electromagnetic field
(the Hojman--Rosenbaum--Ryan--Shepley or HRRS theory of
propagating torsion)~\cite{HRRS1}.\footnote{According to the Palatini principle of minimal coupling, the electromagnetic field
tensor is defined as $F_{\mu\nu}=A_{\nu,\mu}-A_{\mu,\nu}$ and does not
generate torsion~\cite{SL}.}
A generalization of the HRRS theory to non-Abelian gauge fields has been
done in Ref.~\cite{Mu}.

In this paper we point out that the equations of field resulting from
the action with the torsion-modified volume element violate 
positivity of energy for the torsionic scalar.
The solution to this problem is to apply a conformal transformation of the original
metric into a new metric in which the field equations have the form of those
in general relativity~\cite{MS, M}.
This new metric defines the {\em Einstein frame} while the original one
defines the {\em Jordan frame}~\cite{VG}.
In Sec.~2 we review the Einstein--Cartan theory with the torsion-modified 
volume element. 
In Sec.~3 we reformulate this theory in the Einstein conformal frame and
apply it to the gauge invariant electrodynamics in the presence of torsion.
The results are briefly summarized in Sec.~4.

\section{Covariant volume in the presence of torsion}

In general relativity, a minimally coupled theory is constructed by replacing
the metric of special relativity $\eta_{\mu\nu}$ with the metric of general 
relativity $g_{\mu\nu}$, and by replacing ordinary derivatives with covariant 
derivatives (the comma--semicolon rule)~\cite{LL2}.
The covariant derivative of a vector $V_{\mu}$ is defined as
\begin{equation}
V_{\nu:\mu}=V_{\nu,\mu}-\{^{\phantom{\mu}\rho}_{\mu\phantom{\rho}\nu}\}V_{\rho}.
\label{covdev1}
\end{equation}
The coefficients $\{^{\phantom{\mu}\rho}_{\mu\phantom{\rho}\nu}\}$ are the Christoffel
symbols:
\begin{equation}
\{^{\phantom{\mu}\rho}_{\mu\phantom{\rho}\nu}\}=\frac{1}{2}g^{\rho\lambda}(g_{\nu\lambda,\mu}+g_{\mu\lambda,\nu}-g_{\mu\nu,\lambda}),
\label{Chr}
\end{equation}
determined from the relation $g_{\mu\nu:\rho}=0$.
The colon denotes a covariant derivative with respect to the Christoffel symbols, and the comma denotes a usual derivative.
In the presence of torsion 
$S^{\rho}_{\phantom{\rho}\mu\nu}=\Gamma^{\phantom{[\mu}\rho}_{[\mu\phantom{\rho}\nu]}$,
the covariant derivative is given by
\begin{equation}
V_{\nu;\mu}=V_{\nu,\mu}-\Gamma^{\phantom{\mu}\rho}_{\mu\phantom{\rho}\nu}V_{\rho}.
\label{covdev2}
\end{equation}
The semicolon denotes a covariant derivative with respect to the
nonsymmetric connection, and
the connection coefficients $\Gamma^{\phantom{\mu}\rho}_{\mu\phantom{\rho}\nu}$ are now
\begin{equation}
\Gamma^{\phantom{\mu}\rho}_{\mu\phantom{\rho}\nu}=\{^{\phantom{\mu}\rho}_{\mu\phantom{\rho}\nu}\}+S^{\rho}_{\phantom{\rho}\mu\nu}-2S^{\phantom{(\mu\nu)}\rho}_{(\mu\nu)}.
\label{conn}
\end{equation}
This relation results from the metric compatibility of the connection, 
$g_{\mu\nu;\rho}=0$.\footnote{
In the presence of torsion, the relation $g_{\mu\nu:\rho}=0$ remains valid,
which means that both covariant derivatives (with respect to 
$\{^{\phantom{\mu}\rho}_{\mu\phantom{\rho}\nu}\}$ and
$\Gamma^{\phantom{\mu}\rho}_{\mu\phantom{\rho}\nu}$, respectively)
are tensors.}
The difference $\Gamma^{\phantom{\mu}\rho}_{\mu\phantom{\rho}\nu}-\{^{\phantom{\mu}\rho}_{\mu\phantom{\rho}\nu}\}$,
which is a tensor, is called the contortion $K^{\rho}_{\phantom{\rho}\mu\nu}$:
\begin{equation}
K^{\rho}_{\phantom{\rho}\mu\nu}=S^{\rho}_{\phantom{\rho}\mu\nu}-2S^{\phantom{(\mu\nu)}\rho}_{(\mu\nu)}.
\label{cont}
\end{equation}

In a spacetime without torsion, a covariant volume element is $\sqrt{-g}d^{4}x$, 
and the scalar density $\sqrt{-g}$ is connection compatible (parallel):
$(\sqrt{-g})_{:\mu}=0$.
When the spacetime is not torsionless, this element is not parallel since~\cite{Sa3}
\begin{equation}
(\sqrt{-g})_{;\mu}=-2S_{\mu}\sqrt{-g}.
\label{den}
\end{equation}
It is possible to find a parallel volume element if the torsion vector is
the gradient of a scalar~\cite{Sa4, Sa1}:
\begin{equation}
S_{\mu}=\theta_{,\mu}.
\label{covol1}
\end{equation}
In this case we have
\begin{equation}
(e^{2\theta}\sqrt{-g})_{;\mu}=0,
\label{covol2}
\end{equation}
and the volume element becomes
\begin{equation}
dV_{4}=e^{2\theta}\sqrt{-g}d^{4}x.
\label{covol3}
\end{equation}

The contortion tensor~(\ref{cont}) can be split into a traceless part and 
a trace (the torsion vector):
\begin{equation}
K_{\rho\mu\nu}=C_{\rho\mu\nu}-\frac{2}{3}(S_{\rho}g_{\nu\mu}-S_{\nu}g_{\rho\mu}).
\label{split1}
\end{equation}
The Riemann--Cartan curvature tensor is given by
\begin{equation}
R^{\sigma}_{\phantom{\sigma}\mu\rho\nu}(\Gamma)=\Gamma^{\phantom{\nu}\sigma}_{\nu\phantom{\sigma}\mu,\rho}-\Gamma^{\phantom{\rho}\sigma}_{\rho\phantom{\sigma}\mu,\nu}+\Gamma^{\phantom{\nu}\kappa}_{\nu\phantom{\kappa}\mu}\Gamma^{\phantom{\rho}\sigma}_{\rho\phantom{\sigma}\kappa}-\Gamma^{\phantom{\rho}\kappa}_{\rho\phantom{\kappa}\mu}\Gamma^{\phantom{\nu}\sigma}_{\nu\phantom{\sigma}\kappa},
\label{curv1}
\end{equation}
and its contractions are the Ricci tensor $R_{\mu\nu}(\Gamma)=R^{\rho}_{\phantom{\rho}\mu\rho\nu}(\Gamma)$ and the curvature scalar $R(\Gamma,g)=R_{\mu\nu}(\Gamma)g^{\mu\nu}$.
The curvature scalar $R(\Gamma,g)$ can be split into the Riemannian curvature
scalar $R(g)$ (constructed from $\{^{\phantom{\mu}\rho}_{\mu\phantom{\rho}\nu}\}$ the same way as
$R(\Gamma,g)$ is constructed from $\Gamma^{\phantom{\mu}\rho}_{\mu\phantom{\rho}\nu}$) and the
part that contains torsion~\cite{Sa2}:
\begin{equation}
R(\Gamma,g)=R(g)-4S^{\mu}_{\phantom{\mu};\mu}+\frac{16}{3}S_{\mu}S^{\mu}+C_{\mu\nu\rho}C^{\mu\rho\nu}.
\label{decomp}
\end{equation}

The Lagrangian density for the gravitational field in the Einstein--Cartan--Saa
theory with the torsion-modified volume element is given by~\cite{Sa2}
\begin{equation}
{\cal L}=\frac{1}{16\pi}R(\Gamma,g)\sqrt{-g}e^{2\theta},
\label{grav}
\end{equation}
where we use the units in which $c=G=1$.
The second term on the right-hand side of Eq.~(\ref{decomp}) is a total
covariant divergence of a vector
(with respect to the nonsymmetric connection $\Gamma^{\phantom{\mu}\rho}_{\mu\phantom{\rho}\nu}$)
and does not contribute to the field
equations.\footnote{
From Eq.~(\ref{covol2}) we obtain $\int S^{\mu}_{\phantom{\mu};\mu}\sqrt{-g}e^{2\theta}d^4x=\int (S^{\mu}\sqrt{-g}e^{2\theta})_{,\mu}d^4x$.
The last integral can be
transformed into an integral over a three-dimensional hypersurface, which vanishes when we use the principle of least action.}
In the same equation we use the condition~(\ref{covol1}).
For the reason explained in the next section, we replace the scalar field
$\theta$ by another field $\phi$ such that
\begin{equation}
\theta=-\frac{3}{2}\phi.
\label{resc}
\end{equation}
The action for the gravitational field is thus
\begin{equation}
S_g=\int d^{4}x\sqrt{-g}e^{-3\phi}\Bigl(-\frac{1}{16\pi}R(g)-\frac{3}{4\pi}\phi_{,\mu}\phi^{,\mu}-\frac{1}{16\pi}C_{\mu\nu\rho}C^{\mu\rho\nu}\Bigr).
\label{actgrav1}
\end{equation}
We note that if $C_{\mu\nu\rho}=0$, replacing the torsionic scalar by a new
field $\varphi=e^{-3\phi}$ reproduces the Brans--Dicke action with
$\omega=-\frac{4}{3}$~\cite{BD}.

\section{Scalar torsion in the Einstein frame}

In the action~(\ref{actgrav1}), the kinetic term of the torsionic scalar
field has the negative sign.
By a sufficiently rapid change of $\phi$ with time, this term can consequently
be made as large as one likes.
The action would then decrease without limit, that is, there could be no
minimum.
This is a feature of many scalar--tensor theories of gravity in the metric
variational formalism~\cite{M, VG}.
To solve this problem, we need to apply a conformal transformation of
the metric from the original Jordan frame to the Einstein frame.
In the Einstein frame, the curvature scalar in the action is multiplied by a constant only,
and the equations of field have the form of the Einstein equations~\cite{MS}. 
Let us make a conformal transformation from the Jordan metric $g_{\mu\nu}$ to
a new metric $h_{\mu\nu}$:
\begin{equation}
h_{\mu\nu}=e^{\lambda}g_{\mu\nu},
\label{conf1}
\end{equation}
where $\lambda$ is a function of the coordinates.
The Jordan and Einstein curvature scalars are related by~\cite{D}
\begin{equation}
R(g)=e^{\lambda}\Bigl(R(h)+3\lambda^{:\mu}_{\phantom{:\mu}\mu}-\frac{3}{2}\lambda_{,\mu}\lambda^{,\mu}\Bigr),
\label{conf2}
\end{equation}
and all the quantities of the right-hand side of this equation are calculated
using the new metric $h_{\mu\nu}$.

To eliminate the exponential function multiplying $R(g)$ in
Eq.~(\ref{actgrav1}), we put 
\begin{equation}
\lambda=-3\phi.
\label{put}
\end{equation}
The action for the gravitational field in the Einstein frame becomes
\begin{equation}
S_g=\int d^{4}x\sqrt{-h}\Bigl(-\frac{1}{16\pi}R(h)+\frac{3}{32\pi}\phi_{,\mu}\phi^{,\mu}-\frac{1}{16\pi}C_{\mu\nu\rho}C^{\mu\rho\nu}\Bigr),
\label{actgrav2}
\end{equation}
where the traceless part of the contortion in the new action is given
by\footnote{To obtain Eq.~(\ref{actgrav2}), we notice that
we can rewrite~(\ref{split1}) as $K_{\rho\phantom{\mu}\nu}^{\phantom{\rho}\mu}=C_{\rho\phantom{\mu}\nu}^{\phantom{\rho}\mu}-\frac{2}{3}(S_{\rho}\delta_{\nu}^{\mu}-S_{\nu}\delta_{\rho}^{\mu})$ that does not contain the metric, and that $C_{\mu\nu\rho}C^{\mu\rho\nu}$ scales under the conformal transformation~(\ref{conf1}) like $R$.}
\begin{equation}
K_{\rho\mu\nu}=C_{\rho\mu\nu}-\frac{2}{3}(S_{\rho}h_{\nu\mu}-S_{\nu}h_{\rho\mu}).
\label{split2}
\end{equation}
The sign of the kinetic term for the scalar field is now positive.
The action~(\ref{actgrav2}) differs from the expression obtained without
using the parallel volume element~(\ref{covol3}) by the factor $\frac{1}{4}$
in the scalar kinetic term~\cite{Niko}.\footnote{The action in Ref.~\cite{Niko} has
$C_{\mu\nu\rho}=0$.}

Let us apply the obtained results to the case of the electromagnetic field
coupled minimally to torsion.
The electromagnetic field tensor is a spacetime without torsion is given by
\begin{equation}
F_{\mu\nu}=A_{\nu:\mu}-A_{\mu:\nu},
\label{Faraday1}
\end{equation}
and is invariant under a gauge transformation 
$A_{\mu}\rightarrow A_{\mu}'=A_{\mu}+\Lambda_{,\mu}$.
In the presence of torsion, the principle of minimal coupling requires the
following definition:
\begin{equation}
F_{\mu\nu}=A_{\nu;\mu}-A_{\mu;\nu}=A_{\nu,\mu}-A_{\mu,\nu}-2S^{\rho}_{\phantom{\rho}\mu\nu}A_{\rho}.
\label{Faraday2}
\end{equation}
Such a tensor is invariant under a generalized gauge transformation
\begin{equation}
A_{\mu}\rightarrow A_{\mu}'=A_{\mu}+e^{\phi}\delta_{\mu}^{\nu}\Lambda_{,\nu},
\label{pot}
\end{equation}
provided that the torsion tensor is given by~\cite{HRRS1}
\begin{equation}
S^{\rho}_{\phantom{\rho}\mu\nu}=\frac{1}{2}(\delta_{\nu}^{\rho}\phi_{,\mu}-\delta_{\mu}^{\rho}\phi_{,\nu}).
\label{comp}
\end{equation}
This relation means that the torsion tensor is fully determined in terms
of the torsion vector $S_{\mu}=S^{\nu}_{\phantom{\nu}\nu\mu}$, and this vector has a
potential, $S_{\mu}=-\frac{3}{2}\phi_{,\mu}$.
The above constraint on torsion contains the condition for the existence
of a parallel volume element~(\ref{covol1}), but is stronger.
The tensor~(\ref{Faraday2}) becomes
\begin{equation}
F_{\mu\nu}=A_{\nu,\mu}-A_{\mu,\nu}-A_{\nu}\phi_{,\mu}+A_{\mu}\phi_{,\nu},
\label{Faraday3}
\end{equation}
which looks more elegant if we use $\phi$ instead of $\theta$.

The total gauge invariant action for the electromagnetic field and the 
gravitational field with torsion is given by
\begin{equation}
S=\int d^{4}x\sqrt{-h}\Bigl(-\frac{1}{16\pi}R(h)+\frac{3}{32\pi}\phi_{,\mu}\phi^{,\mu}-\frac{3}{32\pi}m^{2}\phi^{2}-\frac{1}{16\pi}F_{\mu\nu}F^{\mu\nu}\Bigr),
\label{act}
\end{equation}
where we introduce a mass of the torsionic field~\cite{Niko, G} (see the next paragraph).
The equations of field are obtained from variation of $h_{\mu\nu}$, $\phi$, 
and $A^{\mu}$:
\begin{eqnarray}
& & G_{\mu\nu}(h)=\frac{3}{2}\Bigl(\phi_{,\mu}\phi_{,\nu}-\frac{1}{2}\phi_{,\rho}\phi^{,\rho}h_{\mu\nu}\Bigr)+\frac{3}{4}m^{2}\phi^{2}h_{\mu\nu} \nonumber \\
& & +2\Bigl(\frac{1}{4}F_{\rho\sigma}F^{\rho\sigma}h_{\mu\nu}-F_{\mu\rho}F_{\nu}^{\,\,\,\rho}\Bigr), \\
& & \phi^{:\mu}_{\phantom{:\mu}\mu}+m^{2}\phi+\frac{4}{3}(F^{\mu\nu}A_{\nu})_{:\mu}=0, \\
& & F^{\mu\nu}_{\phantom{\mu\nu}:\nu}+F^{\mu\nu}\phi_{,\nu}=0,
\label{eof}
\end{eqnarray}
where $G_{\mu\nu}(h)$ is the Einstein tensor.
They differ from the equations derived in the HRRS theory~\cite{HRRS1, Niko} by the factor
$\frac{1}{4}$ in the torsionic scalar field terms.
The last two equations yield
\begin{equation}
\phi^{:\mu}_{\phantom{:\mu}\mu}+m^{2}\phi=-\frac{2}{3}F_{\mu\nu}F^{\mu\nu}.
\label{tors}
\end{equation}

The reason for the torsionic scalar field to be massive originates from the
E\"{o}tv\"{o}s--Dicke--Braginsky solar tests of the principle of equivalence~\cite{N} and the idea introduced in Ref.~\cite{G}.
For the scalar field of the Sun, Eq.~(\ref{tors}) becomes
\begin{equation}
\nabla^{2}\phi-m^{2}\phi=\frac{1}{3}({\bf B}^{2}-{\bf E}^{2}),
\label{weak}
\end{equation}
where ${\bf B}$ and ${\bf E}$ are the magnetic and 
electric field, respectively.
The solution of Eq.~(\ref{weak}) outside the Sun is
\begin{equation}
\phi=\frac{2}{3}\frac{e^{-mr}}{r}E_{ne},
\label{sol}
\end{equation}
where $E_{ne}$ is the total nuclear electric energy of the Sun and
other energies are negligible.
The data from Ref.~\cite{N} give
\begin{equation}
\phi=0.67\times10^{-4}U\cdot e^{-mr},
\label{num}
\end{equation}
where $r$ is the distance from the Sun and $U$ is the Newtonian
potential.
This expression modifies the relative acceleration between aluminum
and gold/platinum~\cite{N} by the
factor $e^{-mR}$, where $R$ is the distance of the Earth from the Sun:
\begin{equation}
{\bf a}_{rel}=2\times10^{-7}\nabla U\cdot e^{-mR}.
\label{exp}
\end{equation}
To avoid violations of the principle of
equivalence in the solar system and obtain a theory
compatible with experiment, we need
\begin{equation}
2\times10^{-7}\nabla U\cdot e^{-mR}<10^{-12}\nabla U.
\label{cons}
\end{equation} 
This inequality gives the lower limit on the mass of the
torsionic scalar field~\cite{Niko},
\begin{equation}
m>10^{-25}\mbox{GeV},
\label{res}
\end{equation}
in agreement with Ref.~\cite{G}.

\section{Summary}

In the presence of torsion, a covariantly conserved volume element can
be found if the torsion vector equals the gradient of a scalar.
This condition gives four equations of constraint on the torsion
tensor.
Remarkably, the condition for the compatibility of the gauge
invariant electromagnetic field coupled minimally to
torsion contains this constraint.
Another condition for this compatibility is that the traceless
part of the contortion tensor must vanish.
The last requirement is a consequence of the equations of field in the Jordan frame, if we use the electromagnetic field that does not couple to
torsion.
It is not the case for the minimal coupling between torsion and photons.
A possible solution would be to assume that the electromagnetic field couples to the
trace part of the torsion tensor only.
Ultimately, we should obtain this requirement as a result of a
variational principle.

The action integral over the torsion-modified volume element contains the kinetic term of the torsionic scalar
field with the negative sign.
By a sufficiently rapid change of this field with time, this term can consequently
be made arbitrarily large, and the action would have no
minimum.
To solve this problem, we applied a conformal transformation of
the metric from the original Jordan frame to the Einstein frame
in which the left side of the field equations is that of
general relativity. 
The new action acquired the correct, positive sign in the kinetic term for
the torsionic scalar.
The obtained field equations differ from those in the original HRRS theory by the factor
$\frac{1}{4}$ in the torsionic scalar field terms.
This difference does not affect the order of the minimal value for the mass $m$ of the torsionic
scalar required for the theory to be compatible with experiment.

The minimal value of $m$ is
way below the masses of known elementary particles.
There is no upper limit for $m$, and the Higgs boson could be a good
candidate for the particle corresponding to the torsionic scalar.
Since the lower limit on the mass of the Higgs boson is on the
order of $100\,\mbox{GeV}$, the deviations from the principle of
equivalence would be unnoticable (below $10^{-27}$ of present
experimental precision).

We emphasize that the question of whether the electromagnetic field
couples to torsion or not should be ultimately answered by
experiment.
We assumed that the principle of minimal coupling holds in the
presence of torsion, which leads to the appearance of the
EM--torsion coupling and constraints on the torsion tensor.
Otherwise, torsion does not affect the electromagnetic field
and gauge invariance of the latter is compatible with nonsymmetric
connection under no additional constraints.

We mentioned that the HRRS procedure of combining electromagnetic
gauge invariance with minimal coupling between the EM field and
torsion has been generalized to non-Abelian gauge fields such
as the Yang--Mills field.
A possible coupling between spin, which generates torsion in
the Einstein--Cartan theory, and the non-Abelian chromomagnetic field of QCD
would play an important role in objects composed of fermionic matter at
large densities.
Neutron stars would be ideal candidates to study this coupling
and determine whether gauge fields interact with torsion.

\end{document}